\documentstyle[12pt]{article}
\begin{document}

\noindent {\large\bf Zero Modes of Rotationally Symmetric Generalized
Vortices and Vortex Scattering}
\vspace{1cm}

\noindent
J. Burzlaff $\left.^{\dag} \right.$\footnote{On leave of absence from the
School of Mathematical Sciences, Dublin City University, Dublin 9, Ireland.}
and D.H. Tchrakian $\left.^{\ddag}\:^* \right.$

\vspace{.7cm}

\noindent $\dag$ Fachbereich Physik, Universit\"{a}t Kaiserslautern, \\
Erwin Schr\"{o}dinger Strasse, D-67653 Kaiserslautern, Germany. \\

\noindent $\ddag$ Department of Mathematical Physics, St Patrick's College, \\
Maynooth, Ireland. \\

\noindent $*$ School of Theoretical Physics,  Dublin Institute  \\
for Advanced Studies, 10 Burlington Road, Dublin 4,  Ireland. \\

\vspace{1.5cm}

\noindent {\large \bf Abstract}
Zero modes of rotationally symmetric vortices in a hierarchy
of generalized Abelian Higgs models are studied. Under the
finite-energy and the smoothness condition,  it is shown, that
in all models, $n$ self-dual vortices superimposed at the origin
have $2n$ modes. The relevance of these modes for vortex
scattering is discussed, first in the context of the slow-motion
approximation. Then a corresponding Cauchy problem
for an all head-on collision of $n$ vortices is formulated.
It is shown that the solution of this Cauchy problem has
a $\frac{\pi}{n}$ symmetry.
\vspace{3cm}

\noindent KL-TH-95/20

\newpage

\noindent {\bf 1. Introduction}

\vspace{.5cm}

Since their discovery \cite{one:1}, vortices in the Abelian Higgs
model have attracted much attention. This is mainly due to the
fact that the static solutions of this model
describe flux tubes in superconductors.
As objects in 2-dimensional space, they also provide simple
examples of topologically nontrivial structures. Recently , the
Abelian Higgs model was generalized, and rotationally
symmetric vortices were found in these models \cite{two:2}. The
{\it generalised} Abelian Higgs models form a hierarchy of $2$
dimensional $U(1)$ models lablled by an integer parameter $p$.
The significance of $p$ is that the $p$-th member of this hierarchy
has been derived by subjecting the $p$-th member of the hierarchy
of $4p$ dimensional scale-invariant $SO(4p)$ Yang-Mills models to
dimensional descent\cite{two-three:2-3}. The $p=1$ members of both
these hierarchies are the usual Abelian Higgs model and the Yang-mills
model in $2$ and $4$ dimensions respectively.

A mathematically rigorous proof for the existence and uniqueness of
the rotationally symmetric self-dual vortices of the hierarchy of
generalised Abelian Higgs models was subsequently given in
\cite{thr:3}. These rotationally symmetric solutions
describe vortices superimposed at the origin, and provide us with
objects qualitatively similar to, but quantitatively different from
the vortices in the Abelian Higgs model. In the case of the
Abelian Higgs model, actually, a $2n$-parameter family of vortices
exists \cite{fou:4}. In this paper, we show that, at least near
the rotationally symmetric vortices, there is also a $2n$-parameter
family of generalized vortices.

In the Abelian Higgs model, the knowledge gained from the
study of the zero modes has been used to investigate the
scattering of vortices. Vortex scattering has been
studied in the context of the slow-motion approximation
\cite{fiv:5}, and with the help of the theory of partial
differential equations \cite{six:6}. Among the most interesting
processes studied is $90^{\circ}$ scattering of $2$ vortices
in a head-on collision (or its generalization: $\frac{\pi}{n}$
scattering in all head-on collisions of $n$ vortices.)
We will show that some of the results, in particular the
$\frac{\pi}{n}$ symmetry of the scattering process, generalize
to generalized vortices.

The paper is organized as follows. In Sec. 2, we highlight those
aspects of the generalized Abelian Higgs models, and of their
rotationally symmetric vortices, important to our discussion of the
zero modes. In Sec. 3, following steps taken by Weinberg
in the case of the Abelian Higgs model \cite{sev:7}, we derive
the fluctuation equations for rotationally symmetric generalized
vortices. We show that these equations have a $2n$-parameter
family of smooth finite-energy solutions. In Sec. 4, the
scattering of $n$ generalized vortices shortly before and
after they form a rotationally symmetric vortex is discussed,
first, briefly, in the context of the slow-motion approximation.
Then a corresponding Cauchy problem is formulated.
It is shown that its solution has a $\frac{\pi}{n}$
symmetry.

\vspace{1cm}

\noindent {\bf 2. The Models and their Radially Symmetric Solutions}

\vspace{.5cm}

The models we study are the generalized Abelian Higgs models
in (2+1)-dimensional space-time, given by the Lagrangian densities,
\begin{eqnarray}
{\cal L}^{(p)} & =  & (\eta^2 - |\phi|^2)^{2(p-2)}
([(\eta^2 - |\phi|^2)F_{\mu \nu} + \imath (p-1)D_{[\mu}\phi^{*}
D_{\nu]} \phi]^2 \nonumber\\[-4mm]
&&\\
&&+ 4p(2p-1)(\eta^2 - |\phi|^2)^2 |D_{\mu}\phi|^2
+ 2(2p-1)^2(\eta^2 - |\phi|^2)^4),\nonumber
\end{eqnarray}
for any integer $p>1$. For $p=1$, (1) reduces to the usual
Abelian Higgs model. $\phi$ is the complex Higgs field,
$D_{\mu}\phi = \partial_{\mu} \phi + \imath A_{\mu} \phi$,
$\mu =0,1,2$, is the covariant derivative, and the gauge fields
$F_{\mu \nu}$ are defined in terms of the real gauge potentials
$A_{\mu}$ as $F_{\mu \nu} = \partial_{\mu} A_{\nu}
- \partial_{\nu} A_{\mu} = \partial_{[\mu} A_{\nu]}$
$\mu, \nu = 0,1,2$. The square brackets mean antisymmetrization
in the indices. For any tensor $T_{\mu \nu ...}$,
$(T_{\mu \nu ...})^2$ means $T_{\mu \nu ...}T^{\mu \nu ...}$.
The indices are raised and lowered with the
metric tensor $g = diag(-1,+1,+1)$. Our task is to find, and
study, smooth finite-energy solutions of the corresponding
Euler-Lagrange equations.

As in the case of the Abelian Higgs model, the
Euler-Lagrange equations of the generalized Abelian Higgs
models are solved by certain first-order
equations. For the Abelian Higgs model, these equations were
found by Bogomol'nyi \cite{eig:8}. For the Lagrangian (1),
the first-order equations arise as follows:
In the case $A_0 = 0$, and for time-independent Higgs field
and gauge potentials, the Lagragian can be written in the form,
${\cal L}^{(p)} = I^2 +J^2 + \epsilon_{ij} \partial^{i} \Omega^{j}$,
where $I^2$ and $J^2$ are the positive definite terms,
\begin{eqnarray}
I^2 & = & (\eta^2 - |\phi|^2)^{2(p-2)} (((\eta^2 - |\phi|^2) F_{ij}
\nonumber\\[-4mm] & & \\
 & & + \imath (p-1) D_{[i}\phi^{*} D_{j]}\phi) - (2p-1) \epsilon_{ij}
(\eta^2 - |\phi|^2)^2)^2, \nonumber\\[3mm]
J^2 & = & 2p(2p-1) (\eta^2 - |\phi|^2)^{2(p-1)}
|D_{i}\phi - \imath \epsilon_{ij} D^{j}\phi |^2,
\end{eqnarray}
and where,

\newpage

\begin{eqnarray}
\Omega^{j} & = & 4(2p-1) \eta^{2(2p-1)} A^{j} \nonumber\\[-3mm]
 & & \\[1mm] & &
+ \sum_{s=1}^{2p-1} \frac{4\imath}{s} (2p-1)^2  \eta^{2(2p-1-s)}
\left( \begin{array}{c}
2p-2 \\ s-1 \end{array} \right)
(-|\phi|^2)^{s-1} \phi^{*} D^{j}\phi.\nonumber
\end{eqnarray}

The special structure of the Lagrangian, ${\cal L}^{(p)}$, is no
accident, but arises naturally through
dimensional reduction of the generalized
Yang-Mills model on $R^2\times S^{4p-2}$ \cite{two:2}. The divergence,
$\epsilon_{ij}\partial^{i}\Omega^{j}$, is the dimensionally reduced
form of the Chern-Pontryagin density. $I=0$ and $J=0$ stem from
the self-duality equations of the generalized Yang-Mills theory
\cite{two-three:2-3}.
Setting $I$ and $J$ equal to zero, means minimizing the energy
in a given tological sector, and yields the desired first-order
equations. We have simplified the task of finding
time-independent solutions with $A_0 = 0$, to solving the following
equations:
\begin{eqnarray}
(\eta^2 - |\phi|^2) F_{ij} + \imath (p-1) D_{[i}\phi^{*} D_{j]}\phi
& = & (2p-1) \epsilon_{ij} (\eta^2 - |\phi|^2)^2, \\
D_{i}\phi & = & \imath \epsilon_{ij} D^{j}\phi.
\end{eqnarray}

For the rotationally symmetric ansatz,
\begin{eqnarray}
\phi & = & \eta g(r) \exp (-\imath n\theta), \\
A_{i} & = & \epsilon_{ij} x^{j} \frac{a(r) - n}{r^2},
\end{eqnarray}
with integer vortex number $n$,
Eqs (5) and (6) reduce to,
\begin{eqnarray}
(1-g^2) \frac{da}{dr} & = & \frac{2}{r} (p-1) a^2 g^2
- \eta^2 (2p-1) r (1-g^2)^2, \\
\frac{dg}{dr} & = & \frac{ag}{r}.
\end{eqnarray}
In Ref. 3, for all $n$, existence and uniqueness of a solution to these
equations with the desired asymptotic behaviour was shown.
For small $r$, $a$ goes to $n$, and $g$ goes like $C_{n}r^{n}$,
where the $C_{n}$ are constants which have to be determined
numerically.  For large $r$, $g$ goes to $1$, and $a/(1-g^2)$
goes like $\sqrt{(2p-1)/(2p)} \eta r$.
In the next section, we will study the zero modes of these
rotationally symmetric solutions.

\vspace{1cm}

\noindent {\bf 3. The Fluctuation Equations and the Zero Modes}

\vspace{.5cm}

We write the fluctuations about the rotationally symmetric
solutions in the following form:
\begin{eqnarray}
\delta \phi & = & \eta g(r) h(r,\theta) \exp (-\imath n \theta), \\
\delta A_1 &= & \frac{1}{r} (-b(r,\theta) \sin \theta
+ c(r,\theta) \cos \theta ),\\
\delta A_2 & = & \frac{1}{r} (b(r,\theta) \cos \theta
+ c(r,\theta) \sin \theta).
\end{eqnarray}
Here, $b$ and $c$ are real functions, and $h=h^1 + \imath h^2$
is a complex function. To count the number of modes, we will
later fix the gauge by setting $h^2$ equal to zero. To discuss
the smoothness of the modes, we will have to use the gauge
freedom and find suitable functions $h^2$.

Equations (5) and (6), linearized in the functions $h$, $b$ and $c$,
then yield the fluctuation equations.
Equation (6) leads to expressions for the functions $b$ and $c$
in terms of the function $h$:
\begin{eqnarray}
b = -r \frac{\partial h^1}{\partial r} - \frac{\partial h^2}{\partial
\theta},\\
c = \frac{\partial h^1}{\partial \theta} - r \frac{\partial h^2}{\partial r}.
\end{eqnarray}
Using these expressions, we obtain from eq. (5) the following
equation for $h$:
\begin{equation}
\frac{1}{r} \frac{\partial}{\partial r} (r\frac{\partial h^1}{\partial r} )
+ \frac{1}{r^2} \frac{\partial^2 h^1}{\partial \theta^2}
- \frac{4(p-1) a g^2}{r(1-g^2)} \frac{\partial h^1}{\partial r}
- g^2 (2(2p-1)\eta^2 + \frac{4(p-1) a^2}{r^2 (1-g^2)^2} ) h^1 = 0.
\end{equation}
Note that $h^2$ does not occur in Eq. (16). Therefore, the
finite-energy solutions of Eq. (16), which lead to smooth fields in a
suitable gauge, yield the zero modes. There are no other modes,
since we can impose the gauge condition, $h^2=0$.

We attempt to solve Eq. (16) in terms of a Fourier series for $h^1$,
\begin{equation}
h^{1}(r,\theta) = \sum_{k=0}^{\infty} (h^{(1)}_{k} (r) \cos k\theta
+ h^{(2)}_{k} (r) \sin k\theta ).
\end{equation}
The Fourier coefficient functions, $h^{(i)}_{k}$, have to satisfy
the equation,
\begin{equation}
\frac{d^2 h^{(i)}_{k}}{dr^2} + ( \frac{1}{r} -
\frac{4(p-1)ag^2}{r(1-g^2)})
 \frac{dh^{(i)}_{k}}{dr} - ( \frac{k^2}{r^2} + 2(2p-1) \eta^2 g^2
+ \frac{4(p-1) a^2 g^2}{r^2 (1-g^2)^2})  h^{(i)}_{k} = 0.
\end{equation}
We want to find all solutions such that, in a suitable gauge, the
rotationally symmetric background fields plus the fluctuations are
$C^{\infty}$ functions on $R^2$ with finite energy.

For small $r$, Eq. (18) reduces to, $r^2 y'' + r y' = k^2 y$,
where $y=h^{(i)}_{k}$. The general solution to this equation is,
$y=c_1 r^k + c_2 r^{-k}$ for $k>0$, and
$y= c_1 + c_2 \ln r$ for $k=0$. The uniqueness of radially
symmetric vortices rules out nontrivial fluctuations for $k=0$.
(We assume here that the uniqueness result, rigorously
proven in Ref. 4 for $p=2$, is in fact true for all $p$.)
For large $r$,
 \begin{equation}
y'' - (p-1) \sqrt{\frac{8(2p-1)}{p}} \eta y'
- \frac{2(2p-1)^2}{p} \eta^2 y = 0,
\end{equation}
holds, with general solution,
\begin{equation}
y = c_3 \exp (\sqrt{\frac{2}{p}} (2p-1)^{\frac{3}{2}} \eta r)
+ c_{4} \exp (- \sqrt{\frac{2(2p-1)}{p}} \eta r).
\end{equation}
Because of the finite-energy
condition, we have to set $c_3=0$. $c_4$ is an arbitrary coefficient.
One can now show that all exponentially decreasing
solutions at infinity lead to solutions at the origin with
$c_2 \neq 0$. Assume the contrary: Then for $c_4 > 0$
there must be a maximum with positive $y$. However,
Eq. (18) shows that, when the first derivative vanishes,
the second derivative is positive. This is a contradiction.
(An analogous argument holds when $c_4$ is negative
and $y$ has to attain a minimum.) Hence, acceptable nontrivial
behaviour at infinity leads to an $r^{-k}$ term at the origin.
For $k>n$ this implies that the energy is infinite.

We are left with the $2n$ modes $h^{(i)}_{k}$ (with all
the other $h^{(j)}_{l}$, for $j\neq i$ or $l\neq k$, equal to
zero) for $0<k\leq n$ and $i=1,2$. Supplemented with
suitable functions $h^2$, these modes lead to the
following smooth finite-energy solutions: For the first
set of $n$ functions, the fluctuations are of the form,
\begin{eqnarray}
\delta \phi = \eta g(r) h^{(1)}_{k} (r) \exp (-\imath (n-k)\theta), \\
\delta A_1 = - (\frac{dh^{(1)}_{k}}{dr}
+ \frac{k}{r} h^{(1)}_{k} ) \sin ((k-1)\theta),\\
\delta A_2 = - (\frac{dh^{(1)}_{k}}{dr}
+ \frac{k}{r} h^{(1)}_{k} ) \cos ((k-1)\theta).
\end{eqnarray}
For the second set of $n$ functions, the fluctuations are of
the form,
\begin{eqnarray}
\delta \phi = -\imath\eta g(r) h^{(2)}_{k} (r) \exp (-\imath (n-k)\theta), \\
\delta A_1 = (\frac{dh^{(2)}_{k}}{dr}
+ \frac{k}{r} h^{(2)}_{k} ) \cos ((k-1)\theta),\\
\delta A_2 = - (\frac{dh^{(2)}_{k}}{dr}
+ \frac{k}{r} h^{(2)}_{k} ) \sin ((k-1)\theta).
\end{eqnarray}

\vspace{1cm}

\noindent {\bf 4. Symmetries and Vortex Scattering}

\vspace{.5cm}

In the Abelian Higgs model, the zero modes can be used to
study vortex scattering in the context of the slow-motion approximation
\cite{nin:9}. The idea of this approximation is that for low
velocities, at each point in time the fields, $A_1$, $A_2$, and $\phi$,
are given by one of the static solutions; i.e., as an ansatz for these
functions one can choose the family of static
solutions after making the parameters time dependent.
$A_0$ is then determined from its equation of motion,
which is of zero order in $t$.
Finally, with this ansatz the action is minimized.

For our purpose,  this idea can be implemented as follows.
Near the rotationally symmetric vortices, we expand the fields,
\begin{equation}
\phi^{i} = \hat{\phi}^{i} + s^{i} (t) \delta\phi^{i}, \;\;\;\;
A^{i} = \hat{A}^{i} + s^{2+i} (t) \delta A^{i}
\end{equation}
Here, $\hat{\phi} = \hat{\phi}^1 + \imath \hat{\phi}^2$ and
$\hat{A}^{i}$ are the the static solutions (7) and (8).
$\delta \phi ^{i}$ and $\delta A^{i}$ are the fluctuations
(21,22,23) or (24,25,26). The equation, which has to
be solved for $A_0$, is the first (for $\nu=0$) of the following
three equations:
\begin{eqnarray}
\partial_{\mu} ((\eta^2 - |\phi|^2)^{2p-3}
((\eta^2 - |\phi|^2) F^{\mu\nu} & + & \imath (p-1)
D^{[\mu}\phi^{*}D^{\nu]}\phi)) + (p-1) \nonumber\\
\times (\eta^2 - |\phi|^2)^{2p-4}(((\eta^2 - |\phi|^2) F^{\mu\nu}
& + & \imath (p-1) D^{[\mu}\phi^{*}D^{\nu]}\phi)
(\phi^{*} D_{\mu}\phi  + \phi D_{\mu}\phi^{*})) \nonumber\\
- \imath p (2p-1) (\eta^2 - |\phi|^2)^{2p-2} & \times &
(\phi D^{\nu} \phi^{*} - \phi^{*} D^{\nu} \phi) = 0 .
\end{eqnarray}
(The other two equations (for $\nu=1,2$) are the second-order
equations of motion for $A_1$ and $A_2$, which we will need later.)

The dynamics, as discribed by the functions $s^{\alpha} (t)$, is
now given by the Lagrange function,
\begin{eqnarray}
L^{(p)} & =  & \int_{R^2} (\eta^2 - |\phi|^2)^{2(p-2)}
(2((\eta^2 - |\phi|^2)F_{0i} + \imath (p-1)D_{[0}\phi^{*}
D_{i]} \phi)^2 \nonumber\\[-4mm]
&&\\
&&+ 4p(2p-1)(\eta^2 - |\phi|^2)^2 |D_{0}\phi|^2) \; d^{2}x.
\nonumber
\end{eqnarray}
The functions $s^{\alpha}$ are found by solving the
Euler-Lagrange equations of this Lagrange function.
Obviously, the slow-motion approximation still leaves us
with very complicated equations to solve.
With our ansatz (27), however, we only attempt to study the
neighbourhood of rotationally symmetric vortices; i.e.,
we can neglegt higher order terms in the functions $s^{\alpha}$.
Even though we are allowed to linearize in $s^{\alpha}$,
we were not able to solve Eq. (28)
for $\nu = 0$.

In the Abelian Higgs model, the
Gauss equation corresponds to Eq. (28) for $\nu = 0$.
Following the same steps we have just discussed for our models,
in the Abelian Higgs model one finds that $A_0 = 0$, and that
the $s^{\alpha}$ are functions linear in time, in the
neighbourhood of the rotationally symmetric vortices.
For the modes with real smooth
$\delta\phi$ this implies $\frac{\pi}{n}$ scattering, if only
times shortly before and shortly after the vortices
coincide are considered. We see that the slow-motion
approximation yields more results in the
Abelian Higgs model. Another difference between
the models discussed here and the Abelian Higgs model
is that in the Abelian Higgs model the validity of the slow-motion
approximation has been proved rigorously \cite{ten:10}.

Instead of pursuing the slow-motion approximation any further,
we will now formulate a corresponding Cauchy problem and
use the symmetries to study the full Euler-Lagrange
equations of the Lagrangian (1). We work with the vector,
$\psi^{T} = (A_0,\partial_{t}A_0,A_1,\partial_{t}A_1,A_2,\partial_{t}A_2,
\phi,\partial_{t}\phi)$, and impose the following initial conditions:
\begin{equation}
\psi (0,\vec{x})^{T} = (\hat{A}_0,0,\hat{A}_1,\delta A_1,
\hat{A}_2,\delta A_2,\hat{\phi},\delta\phi)
\end{equation}
Here, $\hat{\phi}$ and $\hat{A}_{i}$ are the static solutions (7) and
(8). $\delta \phi$ and $\delta A_{i}$ are the fluctuation (21,22,23)
for $k=n$. We concentrate on this type of fluctuations for
simplicity, and because they lead to the interesting $\frac{\pi}{n}$
symmetry in the $n$ vortex scattering process. $\hat{A}_0$ is
the solution of Eq. (28) for $\nu = 0$ at time $t=0$.

The equations of motion we consider are the following
second-order equations: The second-order equation for $A_0$
is, $\partial_{tt} A^0 + \partial_{i}\partial_{t} A^{i} = 0$.
(This equation follows from the Lorentz condition
$\partial_{\mu} A^{\mu} = 0$.) The second-order equations for
$A_1$ and $A_2$ are given by Eq. (28) for $\nu = 1,2$.
The second-order equation for $\phi$ is its Euler-Lagrange
equation,
\begin{eqnarray}
\imath \partial_{\mu} ((\eta^2 - |\phi|^2)^{2p-4}
[(\eta^2 - |\phi|^2) F^{\nu \mu} & + &
\imath (p-1) D^{[\nu} \phi^{*} D^{\mu ]} \phi]
(p-1) D_{\nu} \phi^{*} \nonumber \\
+ p(2p-1) (\eta^2 - |\phi|^2)^2 D^{\mu} \phi^{*})
& + & \frac{1}{2} (p-2) (\eta^2 - |\phi|^2)^{2p-3} \times\nonumber \\
([(\eta^2 - |\phi|^2)F_{\mu \nu}
+ \imath (p-1)D_{[\mu}\phi^{*} D_{\nu]} \phi]^2
& +&4p(2p-1)(\eta^2  -  |\phi|^2)^2 |D_{\mu}\phi|^2 \nonumber \\
+ 2(2p-1)^2(\eta^2 - |\phi|^2)^4) \phi^{*}
& - & \frac{1}{2} (\eta^2 - |\phi|^2)^{2(p-2)}
(((\eta^2 - |\phi|^2) \nonumber \\
\times F^{\mu \nu}
+ \imath (p-1)D^{[\mu}\phi^{*} D^{\nu]} \phi )
& \times &
(F_{\mu\nu} \phi^{*} - (p-1) A_{[\mu} D_{\nu ]} \phi^{*})
\nonumber \\
- 2p(2p-1)(\eta^2 - |\phi|^2) |D_{\mu}\phi|^2 \phi^{*}
& + & \imath p(2p-1) (\eta^2 - |\phi|^2)^2 A^{\mu} D_{\mu} \phi^{*}
\nonumber \\
-2(2p-1)^2(\eta^2 - |\phi|^2)^3 \phi^{*}) & = & 0 .
\end{eqnarray}
The Lorentz condition and Eq. (28) for $\nu = 0$ are considered
as constraints, which, by our choice of initial data, are satisfied
at $t=0$. In the Abelian Higgs model, it has been proved that
the analogous Cauchy problem has a unique solution, and that
the constraints are propagated \cite{six:6}. In the following, we
will assume that also in the present case a unique solution
exists, although a rigorous proof is still missing. (Because of
the complexity of the equations, to rigorously prove existence of
a unique solution seems a very difficult task.)

For a given solution $\psi (t,\vec{x})$, we define the functions,
$\psi_{i} (t,\vec{x}) = M_{i} \psi (t,\vec{x}_{(i)})$ for $i=1,2$,
where $\vec{x}_{(1)}=S\vec{x}$ with
\begin{equation}
S = \left(
\begin{array}{cc}
\cos \frac{2 \pi}{n} & - \sin \frac{2 \pi}{n} \\
\sin \frac{2 \pi}{n} & \cos \frac{2 \pi}{n}
\end{array}
\right),
\end{equation}
and where $\vec{x}_{(2)}^{T}=(x_1,-x_2)^{T}$. The matrices
$M_{i}$ are defined as,
\begin{equation}
M_{1} = \left(
\begin{array}{cccc}
I & 0 & 0 & 0  \\
0 & A & B & 0  \\
0 & -B & A & 0   \\
0 & 0 & 0 & I
\end{array}
\right) \: \: , \: \:
I = \left(
\begin{array}{cc}
1 & 0 \\
0 & 1
\end{array}
\right) \: \: ,
A = \cos \frac{2 \pi}{n} I \:,\:
B = \sin \frac{2 \pi}{n} I ,
\end{equation}
and
\begin{equation}
M_{2} = \left(
\begin{array}{cccc}
-I & 0 & 0 & 0  \\
0 & -I & 0 & 0  \\
0 & 0 &  I & 0   \\
0 & 0 & 0  & C
\end{array}
\right) \: \: , \: \:
CV = V^{*} \; .
\end{equation}

We can now show that, if $\psi (t,\vec{x})$ is a solution of the
Cauchy problem, so are $\psi_{1} (t,\vec{x})$ and $\psi_{2} (t,\vec{x})$.
(To find the symmetries of the initial data $A_0 (0,\vec{x})$, we have
assumed that Eq. (28) for $\nu = 0$ has a unique smooth solution
with asymptotic decay sufficient to satisfy the finite-energy
condition.) The uniqueness of the solution of the Cauchy
problem now implies that, actually, $\psi$, $\psi_{1}$, and $\psi_{2}$
are all the same function. This, in turn, implies that
functions like $|\phi|^2$, $F_{ij}^{2}$, or the energy density
${\cal E}$ are invariant under a $\frac{2\pi}{n}$ rotation, and
under a reflection w.r.t. the $x_1$-axis.
This leads to the following conclusion: If by using functions
like $|\phi|^{2}$,$F_{ij}^{2}$ or ${\cal E}$, there is a way of
defining the positions $(x_{1}^{a}(t),x_{2}^{a}(t)),a=1,...,n$, of exactly
$n$ separate vortices, these $n$ positions must lie on $n$ radial lines
separated by an angle $\frac{2 \pi}{n}$ with equal distance from the
origin. (As in Ref. 7, we can use the the minima of $|\Phi|^{2}$
to define these positions, near the rotationally symmetric vortices.)
Furthermore, one of these radial lines must be the positive $x_{1}$-axis,
or make an angle $\frac{\pi}{n}$ with the positive $x_{1}$-axis. Any vortex
that does not satisfy these conditions immediately leads to $2n-1$ other
vortices, because of the symmetries of our solution. For continuous
solutions, these positions will change continuously such that
at $t=0$ the $n$ positions coincide, and after the collision the vortices
move again on the radial lines just described. Therefore, they can either go
back on the radial lines they came in on, or go back on radial lines shifted
by an angle $\frac{\pi}{n}$. We will study a further symmetry
to show that the second case is realised.

The last transformation we study is
$\vec{x} \rightarrow M\vec{x}$, where $M$ is the orthogonal matrix
\begin{equation}
M = \left(
\begin{array}{cc}
\cos \frac{\pi}{n} & - \sin \frac{\pi}{n} \\
\sin \frac{\pi}{n} & \cos \frac{\pi}{n}
\end{array}
\right).
\end{equation}
Under this transformation the initial data change as follows:
\[
\psi (0,M\vec{x}) = M_{3} \psi (0,\vec{x}),
\]
with
\begin{equation}
M_{3} = \left(
\begin{array}{cccc}
-\sigma & 0 & 0 & 0 \\
0 & C & -D & 0 \\
0 & D & C & 0 \\
0 & 0 & 0 & -\sigma
\end{array}
\right) \: \: , \: \:
\sigma = \left(
\begin{array}{cc}
1 & 0 \\
0 & -1
\end{array}
\right) \: \: ,
C = \cos \frac{\pi}{n} \sigma \:,\:
D = \sin \frac{\pi}{n} \sigma .
\end{equation}
(We have again assumed the uniqueness of the smooth
finite-energy solution, $A_0 (0,\vec{x})$, of Eq. (28) for
$\nu = 0$.) From the uniqueness of the solution of the Cauchy
problem, $\psi (-t,M\vec{x}) = M_{3} \psi (t,\vec{x})$ follows, and we
see that the functions, $|\phi|^2$, $F_{ij}^2$ and ${\cal E}$, are invariant
under the transformation $(t,\vec{x}) \rightarrow (-t,M\vec{x})$.
This establishes $\frac{\pi}{n}$ scattering for $n$ vortices.

\vspace{1cm}

\noindent {\bf 5. Conclusions}

\vspace{.5cm}

The study of vortices in a hierarchy of generalized Abelian Higgs
models, and the comparison with the Abelian Higgs model was continued.
In previous studies we had seen that the rotationally symmetric generalized
vortices are qualitatively similar to, but quantitatively different
from the vortices. Here similar structures were found in the
neighbourhood of the rotationally symmetric vortices. We found
that the zero modes of the generalized vortices have the same
angular dependence as, but radial behaviour different from that
of the zero modes of the vortices in the Abelian Higgs model.

The slow-motion approximation turned out to be of limited value.
(Since the time-independent solutions are not known explicitly
even in the Abelian Higgs model, the slow-motion approximation
is not very successful in this model either.) On the reasonable
assumption that a certain Cauchy problem has a unique solution,
we were, however, able to study the symmetries of certain
solutions. Each solution describes a process where $n$ vortices
approach and form one structure, namely a rotationally
symmetric vortex. From this structure $n$ vortices emerge.
The pattern the vortices create for time $t$ is the same as that
for time $-t$, after a $\frac{\pi}{n}$ rotation.

\vspace{1cm}

{\large Acknowledgements}

\noindent This work was supported in part by the GKSS-Forschungszentrum
Geesthacht and the Human Capital and Mobility grant ERBCHRX-CT93-0632.

\end{document}